# The fundamentals on the non-black black holes

L. Neslušan

*Astronomical Institute, Slovak Academy of Sciences,
05960 Tatranská Lomnica, Slovakia*

**ABSTRACT.** On the basis of general relativity and quantum statistics, it was shown (Neslušan L.: 2009, Phys. Rev. D 80, 024015) that the equation of state (ES) of extremely hot Fermi-Dirac gas in the surface layer of an ultra-relativistic compact object converges to the same form as the relativistic equation of thermodynamical equilibrium (RETE), which is the condition of stability of the object. The description of energy state of a gas particle was completed with the term corresponding with the potential-type energy. The necessity of such the term is set by the demand of convergence of the relativistic particle-impulse distribution law to its Maxwell-Boltzmann form in the classical limit.

The identity of the ES and RETE, both applied to the gas in the object's surface layer, becomes perfect, yielding the stable object, when the object's physical radius is identical to its gravitational radius. In this state, the internal energy of gas particles in a volume of the object's surface layer increases over all limits in the frame of the volume and this opens the question if the horizon of events actually is an insuperable barrier. It seems to be possible that some matter can be temporarily lifted above the surface or, so far, be ejected from the object and can emit a radiation detectable by a distant observer.

In our contribution, we demonstrate a general validity of the functional form of the potential-type energy found in our previous work. The consistency of the RETE with its non-relativistic approximation can occur only for this functional form. We also point out some observational consequences of the approximate identity of ES and RETE before the object collapses, in the proper time, to its gravitational radius as well as the possible observational consequences of the infinitely high internal energy in the surface layer of already collapsed object. In general, we propagate the idea that a lot of phenomena observed at the stellar-sized or supermassive black holes (or not-yet black holes) can be not necessarily related to the structures in a vicinity of the black hole, e.g. to an accretion disk, but they can be linked directly to the behavior of the central, ultra-compact object.

## 1  INTRODUCTION

The black holes are the astrophysical objects situated below the horizon of events. It is believed that nothing, including massless photons, cannot escape from the black hole.

This claim is based on the conclusion by Oppenheimer and Volkoff (1939), who studied the stability of massive compact objects with no internal source of energy, considering a cold degenerated Fermi-Dirac gas, and found that there is no solution for the stable configuration of such the object.

In this contribution, we bring some new evidence and polemize with the above-mentioned traditional claim about the non-existence stability of supercritically massive compact objects as well as with the proof that the horizon of events of such the objects is an insuperable barrier for the macroscopic ejections of matter.

## 2 THE EQUATION OF STATE. NEW THEORY

In our previous paper (Neslušan, 2009; Paper I, hereinafter), we completed the description of the energy state, $W_j$, in the quantum-statistics formula giving the distribution of number density of particles, $\tilde{n}_{gas}$, constituting a relativistic Fermi-Dirac gas in the dependence on the particle impulse, $p$, i.e.

$$\tilde{n}_{gas}(p) = K_N \sum_j \frac{f_j}{\exp\{[W_j(p) - \mu]/(kT)\} + 1}, \tag{1}$$

where $K_N$ is a normalization constant, $k$ is the Boltzmann constant, $T$ is the proper temperature of gas, $f_j$ is the degeneracy of state $j$, and $\mu$ is the chemical potential. Namely, we completed the energy with the term $-W_P$ giving the potential-type energy of a considered particle in the strong gravity of compact object. The energy in Eq.(1) should not include the rest energy, $W_o$, due to the requirement that the impulse distribution must acquire the form of the Maxwell-Boltzmann distribution law in the classical limit. So, the energy in Eq.(1) is

$$W(p) = \sqrt{c^2 p^2 + W_o^2} - W_o - W_P, \tag{2}$$

where $c$ is the velocity of light.

The state quantities of gas, number density $n_{gas}$, pressure $P_{gas}$, and energy density $E_{gas}$, can be calculated as integrals with respect to impulse, $p$, ranging from zero to infinity of integrands $4\pi p^2 \tilde{n}_{gas}(p)$, $4\pi p^3 v \tilde{n}_{gas}(p)$, and $4\pi p^2 W_K \tilde{n}_{gas}(p)$, respectively. $v$ is velocity and $W_K$ is the kinetic energy of a particle. In Paper I, we derived quantum-statistics equation of state (ES) giving the gradient of gas pressure,

$$\frac{dP_{gas}}{dr} = n_{gas}\frac{dW_P}{dr} + (P_{gas} + E_{gas} - n_{gas}W_P - n_{gas}\mu)\frac{1}{T}\frac{dT}{dr}. \tag{3}$$

In a given moment, the total energy, $W_{tot}$, per unit volume of the object's surface layer equals the sum of gas thermal (kinetic-type) and potential-type energies:

$$W_{tot} = E_{gas} + P_{gas} - n_{gas}W_P = sn_{gas}kT + n_{gas}kT - n_{gas}W_P, \tag{4}$$

from which the mean potential-type energy of a particle is

$$-W_P = W_{tot} - (s+1)kT. \tag{5}$$

$s$ is a parameter corresponding to the number of degrees of freedom. Term $n_{gas}kT$ is the thermal energy of pressure.

The chemical potential $\mu$ can be well neglected in the extreme conditions of the ultra-compact objects. After this neglection and taking Eq.(4) into account, the gradient of pressure (3) can be re-written to

$$\frac{dP_{gas}}{dr} = n_{gas}\frac{dW_P}{dr} + n_{gas}\frac{W_{tot}}{T}\frac{dT}{dr}. \tag{6}$$

## 3 THE FUNCTIONAL FORM OF THE POTENTIAL-TYPE ENERGY

In Paper I, we found the potential-type energy of a particle, which was the constituent of the extremely hot Fermi-Dirac gas, in the form

$$-W_P = -W_o \exp(-\nu/2). \tag{7}$$

Unfortunately, the explicit dependence of $W_P$ on the radial coordinate $r$ is not known. On the surface of the ultra-compact object, the components of the metric tensor should be the continuous functions, therefore the component $g_{44}$ (we use also denotation $g_{44} = \exp(\nu)$) must approach the Schwarzschild solution

$$\exp(\nu) = 1 - \frac{R_g}{r} \tag{8}$$

applicable in free space. $R_g$ is the Schwarzschild gravitational radius. We note, if $r \to R_g$, then $\exp(-\nu/2) \to \infty$. It implies that the magnitude $W_P$ of the potential-type energy is very large at the gravitational radius.

Here, we suggest that the formula (7) is more general, valid not only for the extremely-hot-gas particle, but for a particle in arbitrarily curved space-time. It is matter of definition to suppose either the form given by Eq.(7), or to add the constant $W_o$, moreover, i.e.

$$-W_{P2} = W_o - W_o \exp(-\nu/2). \tag{9}$$

In almost flat space characterized with $\exp(\nu) \to 1$, $-W_P$ given by Eq.(7) equals $-W_P \doteq -W_o - Gm_oM/r$. It means that the potential-type energy contains the term corresponding to the rest energy. In this case, the term $-W_o$ in Eq.(2) should be omitted, because it is already subtracted from the energy $W$ within the term $-W_P$. The form (9) for $\exp(\nu) \to 1$ yields the classical, Newtonian potential energy, $-Gm_oM/r$, on the object's surface, where Eq.(8) is valid. $G$ is the gravitational constant, $m_o$ is the rest mass of the particle (related to $W_o$ as $m_o = W_o/c^2$), and $M$ is the total mass of the object.

The functional form (7) or (9) is unique to satisfy the demand of the convergence of the relativistic equation of the thermodynamical equilibrium (RETE) to its classical analogue in the limit of weak field. The RETE was given as (Tolman, 1930a; see also Tolman, 1969)

$$\frac{dP}{dr} = -\frac{E+P}{2}\frac{d\nu}{dr} \tag{10}$$

and its classical analogue is

$$\frac{dP}{dr} = n\frac{dW_P}{dr} = -n\frac{Gm_oM}{r^2} \tag{11}$$

in terms of number density, $n$, and potential energy, $-W_P$.

The proof of the convergence can be done with the help of the Tolman-Ehrenfest (Tolman, 1930b; Tolman & Ehrenfest, 1930) relation that the product of the proper temperature of material fluid, $T$, and square root of the component $g_{44}$ of metric tensor is a finite constant (we denote this constant by $T_c$). Or the relation can be re-written as

$$T = T_c \exp(-\nu/2). \tag{12}$$

Using this relation and Eq.(7) or (9), we can show that

$$\frac{dT}{dr} = \frac{T_c}{W_o} \frac{dW_P}{dr}. \tag{13}$$

Since $E_{gas} = s n_{gas} kT$ and $P_{gas} = n_{gas} kT$, the RETE (see Eq.(10)) for gas (radiation can be neglected in the classical limit) becomes

$$\frac{dP_{gas}}{dr} = -\frac{1}{2}(s+1)n_{gas}kT\frac{d\nu}{dr} = (s+1)n_{gas}k\frac{dT}{dr}. \tag{14}$$

Replacing the derivative $dT/dr$ according to Eq.(13), the RETE in form (14) actually changes to the classical equation of thermodynamical equilibrium (11). The last condition of this identity is the validity of relation

$$W_o = (s+1)kT_c, \tag{15}$$

which can be confirmed if the classical limit of the RETE (10) for gas is found in the following alternative way (using Eq.(12)). We can write

$$\frac{dP_{gas}}{dr} = -\frac{1}{2}(s+1)n_{gas}kT_c \exp(-\nu/2)\frac{d\nu}{dr} =$$
$$= (s+1)n_{gas}kT_c \frac{d\exp(-\nu/2)}{dr}. \tag{16}$$

At the surface, Eq.(8) is valid, therefore

$$\frac{d\exp(-\nu/2)}{dr} = -\frac{1}{2}\frac{R_g}{(1-R_g/r)^{3/2}r^2} = -\frac{GM}{c^2 r^2}, \tag{17}$$

where we neglected ratio $R_g/r$ in the last equation, since $r \gg R_g$ in the classical limit. Using this expression, we can re-write Eq.(16) to the form

$$\frac{dP_{gas}}{dr} = -(s+1)n_{gas}kT_c\frac{GM}{c^2 r^2}. \tag{18}$$

The right-hand side must be equal to $-G n_{gas} m_o M / r^2$ at the same time. It is possible only if Eq.(15) is valid.

Remark: In the distant outer space, the signal carrying the information about the mass of the ultra-compact object is red-shifted about the factor $\exp(\nu/2)$. In other words, the potential energy of a distant particle in the force field of the object is red-shifted about this factor. The mass and corresponding force must, however, be always conserved. So, it must also be conserved during the collapse of the object, i.e. during the change of $g_{44}$. This conservation is possible only in the case of the functional form of $-W_P$ given by Eq.(7) or Eq.(9). In the place of the distant particle (distant observer), the red-shift factor $\exp(\nu/2)$ is vanished by $\exp(-\nu/2)$ in Eq.(7) or Eq.(9), therefore the outer observer measures the same potential energy as in the non-relativistic case, without any red-shift. For other than $\exp(-\nu/2)$ functional form of $W_P$, the distant observer would detect an unacceptable change of the object's mass during the change of $g_{44}$. The gravity of the object would unacceptably either increase or decrease during this change.

## 4 THE STABLE CONFIGURATION OF THE ULTRA-COMPACT OBJECT

The stable configuration of the object is established, when the gradient of the total pressure, $dP/dr$, satisfies the RETE, i.e. Eq.(10). Inside the hot, compact object, the total pressure consists of two partial pressures: gas pressure and pressure of radiation. Let us now to deal with the gradient of the gas pressure.

Using relations $E_{gas} = sn_{gas}kT$ and $P_{gas} = n_{gas}kT$, as well as Eqs.(12), (15), (7) or (9), the right-hand side of the RETE (10) for the gas can be given as

$$\frac{dP_{gas}}{dr} = -\frac{1}{2}(s+1)n_{gas}kT\frac{d\nu}{dr} = n_{gas}\frac{dW_P}{dr}. \qquad (19)$$

Comparing this to the ES (6), both forms are identical when $W_{tot} = 0$.

As stated in Paper I, the internal energy of the compact-object surface layer per particle is much larger than the rest energy of particle. A long time before collapse, this energy is, on contrary, much smaller than $W_o$, therefore one can expect that $W_{tot} \ll W_o$. The conservation of the energy requires the conservation of this sharp inequality. It means, energy $W_{tot}$ becomes more and more negligible as the physical radius of the collapsing object approaches $R_g$. It can transparently be seen if we re-write Eq.(6), using Eqs.(12), (15), and (7) or (9), to form

$$\frac{dP_{gas}}{dr} = n_{gas}\frac{dW_P}{dr}\left[1 + \frac{W_{tot}}{W_o}\exp(\nu/2)\right]. \qquad (20)$$

When the physical radius of the object $R_b \to R_g$, then $\exp(\nu/2) \to 0$, therefore the second term in the brackets vanishes and ES (20) becomes identical to the RETE in the form (19).

Inside the object with an extremely high internal energy of constituting gas, an intensive radiation must occur. In Paper I, we proved that the state quantities would diverge if the gas consisted of non-zero-rest-mass Bose-Einstein particles. For the zero-rest-mass photons (and, then, $W_P = 0$), the pressure of the intensive radiation can be well described by the well-known relation $P_{rad} = aT^4/3$ being valid regardless an external force is acting or not. In the latter, $a$ is the radiation constant. At the same time, the energy density of the radiation can be given as $E_{rad} = 3P_{rad}$. Considering these relations and Tolman-Ehrenfest relation (12), we can prove that the condition (10) is also satisfied for the radiation.

For $R_b = R_g$, the object disappears below the event horizon and no cooling (except of some energy fluctuations leading to occasional mass ejections; see below) is possible. Summarizing the above considerations, the inevitable consequence of no cooling and identity of the ES and RETE is the permanent stability of the object.

# 5 THE CONCLUSION AND POINTING OUT SOME OBSERVATIONAL CONSEQUENCES

The relativistic, ultra-compact object becomes more and more stable, when its proper physical radius approaches the gravitational radius. It means that the final stage of the collapse of the object below the event horizon is slow.

If the ultra-compact object has accumulated some internal energy in its previous evolution ($W_{tot} > 0$ in Eq.(20)), the gradient of the internal pressure is slightly larger than the gravity, which forces the object to shrink. Until this internal energy is lost, the object will not collapse to the proper physical radius identical with its gravitational radius. Such the object can be regarded as a "not-yet black hole". In Paper I, we demonstrated that the time scale of the energy loss is proportional to the mass of object. It implies that especially among very massive objects in the centers of galaxies and quasars, there could be a lot of not-yet black holes. Some phenomena observed at the black-hole candidates, quasars, active galactic nuclei, jets from galactic centers, etc. can be influenced by an activity of such the not-yet central black hole.

However, even if the compact object collapsed below its event horizon, its interaction with its neighbourhood, also other than the gravitational attraction, cannot be excluded. In Paper I, it was shown that the density of internal energy of the compact-object surface layer increases above all limits when its proper radius $R_b \to R_g$. This infinite energy should be sufficient a matter could escape from the object. Thus, the horizon of events does not longer seem to be any insuperable barrier for an escape of macroscopic volumes of matter from the object due to, e.g., some energy fluctuations. A matter that eventually escapes contains an extreme internal energy which is fastly lost in the outer space via, likely, an intensive, very energetic radiation. Unless the ejected clump is kept together by an external force (e.g. strong magnetic field), it speedly expands. The possibility of the macroscopic ejections from the ultra-compact object opens the question, for example, whether the jets escaping from some galactic centers do originate from the accretion disks, which are assumed to exist around the central compact object, or these jets originate directly from this object.

If the surface of the object, which already collapsed below the event horizon, is not perfectly smooth, the spikes of surface waves can probably occur above the horizon during a long period and, eventually, emit a largely red-shifted radiation, the detection of which cannot be excluded. In the light of this possibility, it is not sure whether all the radiation detected at the black-hole candidates comes from a matter in the vicinity of the candidate (from an accretion disk) or at least a part of it comes directly from the central object.

We believe that the presented new light on the physics of black holes, as well as their preceding (not-yet black-hole) stage, will be helpful in observation of these objects and, especially, at the subsequent representation of the observational data.